\documentstyle[twocolumn,epsfig]{mn2e}
\newcommand{\mincir}{\raise
-2.truept\hbox{\rlap{\hbox{$\sim$}}\raise5.truept
\hbox{$<$}\ }}
\newcommand{\magcir}{\raise
-2.truept\hbox{\rlap{\hbox{$\sim$}}\raise5.truept
\hbox{$>$}\ }}
\newcommand{\minmag}{\raise-2.truept\hbox{\rlap{\hbox{$<$}}\raise
6.truept\hbox{$>$}\ }}
\newcommand{\be}{\begin{equation}}
\newcommand{\ee}{\end{equation}}
\newcommand{\ba}{\begin{eqnarray}}
\newcommand{\ea}{\end{eqnarray}}
\newcommand{\brr}{\begin{array}}

\newcommand{\err}{\end{array}}
\newcommand{\bc}{\begin{center}}
\newcommand{\ec}{\end{center}}

\title{Ages of Elliptical Galaxies: Single versus Multi Population Interpretation}
\author[T. P.\ Idiart, J. Silk and J. A.\ de Freitas Pacheco]
{T.P.\ Idiart$^1$, J. Silk$^2$, Jos\'e A.\ de Freitas Pacheco$^3$\\
\\
$^1$ Instituto de Astronomia, Geof\'isica e Ci\^encias Atmosf\'ericas,
Rua do Mat\~ao 1226, S. Paulo, Brasil\\
$^2$Astrophysics, Denys Wilkinson Building, Keble Road, Oxford OX1 3RH, UK\\
$^3$Observatoire de la C\^ote d'Azur, B.P.4229, F-06304 Nice Cedex 4, France \\
emails: thais@astro.iag.usp.br; silk@astro.ox.ac.uk; pacheco@obs-nice.fr
}
\date{august/2007}

\begin{document}

\maketitle

\begin{abstract}

New calibrations of spectrophotometric indices of elliptical
galaxies as functions of spectrophotometric indices are presented, permitting
estimates of mean stellar population ages and metallicities. These calibrations are 
based on evolutionary models including a two-phase interstellar medium, infall and a galactic wind.
Free parameters were fixed by requiring that models reproduce the mean
trend of data in the color-magnitude diagram as well as in the plane of indices
$H\beta$-$Mg_2$ and $Mg_2$-$<Fe>$. To improve the location of faint ellipticals
($M_B > -20$) in the $H\beta$-$Mg_2$ diagram, down-sizing was introduced.
An application of our calibrations to a sample of ellipticals and a
comparison with results derived from single stellar population models
is given. Our models indicate that mean population ages span an interval
of 7-12 Gyr and are correlated with metallicities, which range
from $\sim$ half up to three times solar. 
  
\end{abstract}


\section{Introduction}

The formation of galaxies is certainly one of the most challenging open problems in cosmology, since
theory must account for the evolution and integrated galaxy properties (Silk 2004). Among
the different morphological types, elliptical (E) galaxies are the simplest ones, with mass indicators
like the central velocity dispersion presenting robust correlations with different 
spectrophotometric properties (see de Freitas Pacheco, Michard \& Mohayaee 2003 for a review).
From a theoretical point of view, different formation scenarios have been proposed: one is monolithic
collapse, in which the gaseous material is assembled either in the form of a unique cloud 
(Larson 1974a) or by interaction and merging of a hierarchy of many primeval lumps of matter, not
including pre-existing stars, into a proto-galaxy (Peebles 2002; Matteucci 2003). The alternative picture, 
that of hierarchical merging, considers that galaxies form from successive non-dissipative mergers of
small-scale halos over a wide redshift range (White \& Rees 1978; Toomre 1977; Kauffmann 1996;
Baugh et al. 1996, 1998).  Thus, a fundamental aspect is whether any of the observable 
properties of Es contains an imprint of their
previous formation history which, in essence, is related to the stellar population formation history.
Therefore, the key questions are when the bulk of stars was formed and if ellipticals evolved passively or were
modified, and to what extent, by interactions with the environment.

In this context, the age distribution of stellar populations in E galaxies is essential to
understand their origin and evolution. Absorption line indices combined with colors are powerful tools 
to derive ages and metallicities of the stellar populations constituting early type galaxies. However, the 
interpretation of these integrated spectral characteristics necessarily requires the use of models. In the 
past, this exercise has been accomplished by different authors using either single stellar population 
models (SSP) or evolutionary models (EVM).

SSP models are straightforward applications of the theory of stellar evolution. Given an initial
mass function (IMF) and chemical composition, the evolution of such a population and its corresponding 
spectral characteristics are completely defined. Properties of E galaxies
derived from observed spectral indices and SSP models were obtained by Worthey (1994), Trager et al.(2000), 
Thomas, Maraston \& Bender (2003) and, more recently by Howell (2005), among
others. Most of these studies have concluded that E galaxies present a small spread in metallicity
but span a wide range of ages.

Elliptical galaxies are certainly not single population systems and the reason generally invoked to interpret
data by using SSP models, besides simplicity, is that the bulk of stars in Es  formed
on a short time-scale and, consequently all stars should have similar ages. Even if the age range of the population mix
constituting the galaxy would be quite narrow, the
build-up of chemical elements requires successive stellar generations, which are necessarily described by
a metallicity distribution. Analyses of mid-ultraviolet indices by Lotz, Ferguson \& Bohlin (2000)
clearly demonstrate that single-age and single-metallicity populations are able to explain
globular cluster data (true single population systems) but not Es. Analytical 
models are, in general, of the ``one-zone" type 
including (or not) mass loss from galactic winds or infall of matter from the intergalactic 
medium. Star formation begins when a critical gas density is attained and the evolution of 
the population mix is calculated by summing the properties of SSP models, representative of 
successive stellar generations. Their formation rate is
controlled by the amount of  residual gas available for star formation, whereas their chemical composition
results from the progressive enrichment in ``metals" of the interstellar medium (ISM) by matter ejected
from stars, notably supernovae. Examples of evolutionary models can be found in Matteucci \& Tornamb\`e (1987),
Bressan, Chiosi \& Tantalo (1996), Vazdekis et al. (1996), Kodama \& Arimoto (1997), Idiart, Michard 
\& de Freitas Pacheco (2003, hereafter IMP03). In general, these
models lead to mean metallicities and $[Mg/Fe]$ ratios comparable to those derived from SSP models, but 
significant disagreements  remain on age determinations of the bulk stellar population. 

In the present work, new calibrations of spectrophotometric indices permitting estimates of mean ages
and metallicities of the stellar population mix constituting early-type galaxies are reported. These
calibrations are based on evolutionary models in which the free parameters were adjusted
in order to  adequately reproduce not only the color-magnitude diagram (CMD) but also the observed strength of
indices like $H\beta$, $Mg_2$ and $<Fe>$. In spite of adopting the classical "one-zone" approximation,
the present model simulates the existence of a two-phase ISM in which stars are formed
only in cold gas regions, allowing also a gradual assembly of the galaxy as well as
the presence of a galactic wind. The anti-correlation between the indices $H\beta$ and $Mg_2$ 
can be explained if we make appeal to the down-sizing effect (Cowie et al. 1996). Thus,
in our model sequence, the less massive galaxies are assembled later ($z \sim$ 0.8) than the more 
massive ones ($z \sim$ 3-4). We have also assumed that dust is mixed with the residual gas producing
a small internal reddening, which improves the fit of the aforementioned diagrams. This paper is organized
as follows: in Section 2 the main features of the model are described; in Section 3 the results for
a grid of fiducial models are presented as well as the resulting calibrations for mean ages and
metallicities; finally, in Section 4 the main conclusions are given.

\section{The Model}

\subsection{The mass balance equations}

The model is based on the ``one-zone" approximation and thus it cannot predict spatial
variations. However, when performing the mass balance, we have assumed that
the gas is either in a hot or in a cold phase, where the former is a consequence of
mass ejection by evolved stars and  hot gas ``cavities"
produced by supernova explosions. The presence of hot gas in bright ellipticals is well established 
by X-ray observations,
which indicate the existence of diffuse thermal  emission in these objects and,   
observations indicate also that cold molecular clouds and not the hot gas phase 
are the sites of star formation. Two-phase models have been considered in the past and,
among others, we mention the work by Ferrini \& Poggianti (1993) and that by Fujita, Fukumoto \& Okoshi (1996), who
have initially adopted an ``one-zone" model and, in a subsequent investigation, considered a spherical
galaxy with a given mass distribution to study local properties (Fujita, Fukumoto \& Okoshi 1997). 
Here, a more simplified version of these approaches is considered. In order 
to establish the mass balance equations, we consider that, as a consequence of the stellar evolution, the gas 
returns to the medium, contributing to the hot phase. Supernovae are additional sources 
of hot gas since they inject mechanical
energy into the interstellar medium through blast waves produced by the explosion. The onset of a galactic
wind removes hot gas from the system and radiative cooling followed by recombination processes
transforms part of the hot gas into the cold phase.

Define the gas fraction $f_g(t)$ as the ratio between the total gas mass present in the galaxy at
instant $t$ and $M_0$, a quantity
representing the
total mass acquired by the galaxy by ``infall" processes (``Infall" here means continuous accretion or
sudden variations of mass by merging events). If $f_h(t)$ and $f_c(t)$ are respectively the gas fractions
in the hot and in the cold phases, define $x_h(t) = f_h(t)/f_g(t)$ and $x_c(t) = f_c(t)/f_g(t)$ such that
$x_h(t)+x_c(t) = 1$. Under these conditions, two equations are required to describe the evolution of the
total gas fraction $f_g(t)$ and, for instance, the fraction $x_h(t)$ in the hot phase, since 
$x_c(t) = 1 - x_h(t)$. 

The equation governing the total gas fraction $f_g(t)$ evolution can be written as
\begin{eqnarray}
\frac{df_g(t)}{dt} = -k(1-x_h(t))f_g(t)+\frac{df_{ej}(t)}{dt}\nonumber
\\
-\frac{x_h(t)f_g(t)}{\tau_w}+\dot R_{in}
\end{eqnarray}
The first term on the right represents the amount of cold gas which is transformed into stars. The star
formation rate normalized with respect to $M_0$, $R_*(t)=kx_c(t)f_g(t)$, was assumed to 
be proportional to the available amount 
of cold gas, with an efficiency $k$ given in Gyr$^{-1}$. The second term represents the gas 
returned to the ISM at a rate
\begin{equation}
\frac{df_{ej}(t)}{dt} = k\int^{m_{s}}_{m(t)}(m-m_r)\zeta(m)R_*(t-\tau_m)dm
\end{equation}
The upper limit of the integral was taken equal to $80 M_{\odot}$ and the lower limit corresponds to the stellar
mass whose lifetime is equal to $t$. In the integrand, $m_r$ is the mass of the remnant, which depends on the
progenitor mass, $\zeta(m) = A/m^{\gamma}$ is the initial mass function (IMF)
and the star formation rate is to be taken at the retarded time $(t-\tau_m)$, where $\tau_m$ is lifetime of
a star of mass $m$. The adoption of a power-law at low masses for the IMF
is a simplifying assumption that has no bearing on our results, which depend only on the slope at the massive 
end, which we indeed will vary as a model parameter. The third term represents the mass loss by the galactic 
wind, whose rate was assumed to
be proportional to the amount of hot gas. Finally, the last term gives the rate at which the galaxy accretes
mass, here assumed to be of the form $\dot R_{in} \propto e^{-t/t_{in}}$.

The second equation, which describes the evolution of the fraction $x_c(t)$ of hot gas is 
\begin{eqnarray}
\frac{dx_h(t)}{dt} = -x_h(t)\frac{dlgf_g(t)}{dt}+\frac{m_H}{M_0}
\frac{1}{f_g(t)}\sum Q_i\nu_i\nonumber
\\
+\frac{1}{f_g(t)}\frac{df_{ej}}{dt}
-\frac{\alpha(T)M_0}{m_HV_g}x_h^2(t)f_g(t)-\frac{x_h(t)}{\tau_w}
\end{eqnarray}
In this equation, the first term on the right, since $x_h(t)$ is defined with respect to the total gas fraction, takes 
into account the time variation of the ``background". The second term represents 
the production of hot gas by supernovae
and the sum is performed over both types Ia and II; $Q$ is the mean number of atoms converted into the hot phase per
explosion. This number corresponds to the amount of interstellar gas swept by the blast wave when
forming the ``hot cavity". Using 
the Sedov theory, $Q = 1.35E/(m_HV_s^2)$, where $E$ is the energy of the explosion and $V_s$ is
the shock velocity. As in Fujita, Fukumoto \& Okoshi (1997), we assume that the hot gas cavity ends its
evolution when the shock velocity is comparable to the stellar velocity dispersion. 
The frequency $\nu_{II}$ of type II supernovae was estimated by considering that only
stars in the range $9-45 M_{\odot}$ explode, whereas the frequency $\nu_{Ia}$ of type Ia supernovae was calculated as
in Idiart, de Freitas Pacheco \& Costa (1996). The third term corresponds 
to the contribution of the returned gas (see above) and
the fourth term gives the rate at which the hot gas, after cooling, recombines into neutral and cold gas. In this
term, $\alpha(T)$ is the effective recombination coefficient and $V_g$ is the volume occupied by the gas, identified
with the volume of the galaxy. We have assumed a spherical galaxy whose radius is fixed at any time by the
radius-mass relation given by Gibson (1997). Finally, the last term gives the rate at which the  
hot gas is removed by the galactic wind.
 
Besides the gas evolution, the chemical enrichment of the galaxy is also followed. The chemical evolution 
is described by the usual equations but includes the contribution of the "infall" term, which
dilutes the abundances of heavy elements and the contribution of the wind, which removes enriched gas and modifies
the chemical composition of the nearby primordial intergalactic medium. Yields for massive stars were taken
from Nomoto et al. (1997a) and those from type Ia supernova were taken from model W7 by Nomoto et al.
(1997b). 

\subsection{Spectrophotometric indices}

Integrated luminosities for SSP models for different photometric filters were 
recalculated by taking into account variations in the exponent of the IMF,
in order to obtain self-consistent results. In fact, as in our previous study 
(IMP03), small variations of the IMF exponent $\gamma$ were required 
in order to increase the magnesium yield. The color-magnitude diagram (CMD) of ellipticals
is conventionally interpreted as a mass-metallicity sequence. Larson (1974b) suggested that such a
sequence could be explained by a galactic wind able to gradually halt the star formation process
as the galaxy mass increases. However, a longer duration of the star formation activity also increases
the contribution of type Ia supernovae to the enrichment of the ISM, leading to
an enhancement of iron relative to $\alpha$-elements , contrary to the observed trend (Matteucci \&
Tornamb\`e 1987; Worthey 1998). This problem can be partially surmounted if one assumes that massive
E galaxies have flatter IMFs (Worthey 1998; Tantalo et al. 1998). In fact, in spite
of all uncertainties involving theoretical yields from type II supernovae, there is agreement between
different calculations that the $Mg/Fe$ ratio increases considerably with decreasing metallicity of
progenitors. Moreover, such a ratio varies strongly according to the progenitor mass: the ejecta of 
type II supernova progenitors with masses around 10-13 M$_{\odot}$ have an abundance ratio 
$[Mg/Fe] \sim$ -1.9. This ratio is near solar for
progenitors of $\sim$ 20 M$_{\odot}$ and suprasolar ($[Mg/Fe] \sim$ +1.4) for masses around 70 M$_{\odot}$
(Thielemann et al. 1996). Clearly,
a flatter IMF for increasing galaxian masses alleviates not only the iron index problem but also
permits, in combination with a varying star formation efficiency, a better representation of the
CMD.

Evolutionary tracks required for our computations of photometric indices were taken from 
Girardi et al. (2000) and Salasnich et al. (2000), who have computed stellar models 
with non-solar abundances. The conversion of the iron abundance to the metallicity
Z was performed by using the relation by Salaris et al. (1993), parameterized as a function of the excess of
$\alpha$-elements with respect to solar values and corrections suggested by Kim et al. (2002) for
high metallicities. As first pointed out by Borges et al. (1995), population synthesis based on empirical
stellar libraries inevitably reflects the chemical history of our Galaxy. Thus, functions describing
the variation of indices in terms of atmospheric parameters like temperature, gravity and [Fe/H] must
also include the dependence on the $[\alpha/Fe]$ ratio. We have also revised our previous index
calibrations (Idiart \& de Freitas Pacheco 1995; Borges et al. 1995), including additional stellar data to
our library in order to have a better representation of solar and non-solar objects\footnote{The required
fitting functions for single population synthesis are available under request to the authors}.

\subsection{The computation procedure}

In IMP03, models were tailored to reproduce the CMD of Es in Coma and Virgo, since this relation is
tight for cluster ellipticals, whereas the scatter increases for Es in the nearby field and in small
groups (Schweizer \& Seitzer 1992). Here a different approach was adopted. Since we have searched
for models able to reproduce simultaneously different spectrophotometric indices, a sample of 
E-galaxies was prepared,
including objects with available total magnitudes, which are more representative of colors computed by 
modelers (see, for instance, a discussion in Scodeggio 2001 and Kaviraj et al. 2005) as well as
other spectrophotometric data ($H\beta$, $Mg_2$ and $<Fe>$  indices), required to fix the parameters
of our models. 

The system of equations to be solved has three main free parameters: the star formation efficiency $k$, the
``infall" time-scale and the exponent of the IMF. The wind time-scale $\tau_w$ is not a real independent
parameter as we shall see below. Initially, a grid of models was computed under the assumption that the star 
formation activity begun at the same time for galaxies of all masses. If this hypothesis implies that
galaxies have the same age, it does not imply that their stellar populations have the same mean age, because
the star formation rate is not necessarily the same for galaxies of different masses. Each model in the grid
is characterized by the total mass $M_0$ acquired by infall. Once this parameter is fixed, for a given value 
of $t_{in}$, the initial accretion rate is fixed. In a second step, we adopt a value for the star formation 
efficiency $k$. Miranda (1992) derived from hydrodynamic models for the spherical collapse, including a 
galactic wind, a relation between the 
fraction of gas lost $f_w$ by the galaxy and parameters characterizing the gravitational potential well and
the star formation efficiency, as an indicator of the supernova heating. His results can be quite well fitted by
the expression
\begin{equation}
log f_w = 2.150 - 0.294~log~M_0 + 0.222~log~(kT_{age})
\end{equation}
Adopting this relation, the value of the wind parameter $\tau_w$ is immediately fixed. The  numerical
solution was performed without using the ``instantaneous recycling approximation", leading
to the metallicity distribution function required to compute the integrated spectrophotometric
properties as in Idiart, de Freitas Pacheco \& Costa (1996). For different values of
$M_0$, a sequence of models was computed in which the aforementioned free parameters were 
varied in order to reproduce data on the diagrams $(U-V)-M_B$, $H\beta - Mg_2$ and $Mg_2 - <Fe>$. 

If, as in IMP03, the trend of data in the CMD can
be well reproduced, this approach leads to faint galaxies with 
$H\beta$ indices systematically smaller than observations.
A possible interpretation of the observed inverse correlation between the indices $H\beta$ and $Mg_2$ was
already given in the early 90s either by Sadler (1992) or by Faber, Worthey \& Gonzalez (1992), who
concluded that the stellar populations of the brightest Es are older than less luminous objects. This
is consistent with the work by Cowie et al (1996), who,  based on analysis of the luminosity
function in the K band of deep fields, concluded that ``the mass of star forming galaxies decreases with
redshift",  immortalized as the down-sizing effect (see also Juneau et al. 2005). More recent data derived from deep surveys
($z \sim 1-2$) reveals an excess of massive galaxies with respect to predictions of the 
hierarchical scenario (Cimatti et al. 2004; Glazebrook et al. 2004) or, in
other words, that massive ellipticals were
already in place at these high redshifts with little subsequent merging, whereas less massive Es
present features characteristic of star formation in the more recent past (Ferreras \& Silk 2000). 
All these observations are contrary to expectations for the hierarchical growth of structure in a 
cold dark matter-dominated universe, in which large halos form late by coalescence of smaller ones.

These facts lead us to consider models with an ``inverted hierarchy" by assuming that less massive
galaxies are assembled later than massive ones and, as a consequence, being constituted by a
stellar population mix relatively younger than that of bright galaxies. In practise, this
means we have abandoned the previous idea in which the starting points for mass assembly and
for the beginning of the star formation activity are the same whatever the mass of the galaxy.
The departure time for assembly and the beginning of star formation activity was then adjusted
in order to improve the location of modeled galaxies in the $H\beta-Mg_2$ diagram. 

Moreover, observations suggest that some dust may be mixed with the residual 
gas in E's (e.g. Leeuw et al. 2004; Temi, Brighenti \& Mathews 2007).
A small internal extinction will in particular affect the bluer colors, requiring a slightly higher star
formation efficiency to compensate for such a reddening and produce an increase of the $H\beta$ strength.
This effect goes in the same sense as down-sizing. Thus,
we have included this possibility in our models by assuming that dust is homogeneously mixed with
stars. The transfer equation was solved by considering a plane-parallel geometry and, under these
conditions, the reddened magnitude in a given filter is given by
\begin{eqnarray}
M_{\lambda} = M_{\lambda,0} - 2.5~log~\lbrack\frac{1}{\tau_{eff}}\frac{1}
{(\sqrt{1-\omega_{\lambda}}+coth~\tau_{eff})}\rbrack \nonumber
\\
= M_{\lambda,0} + A_{\lambda}
\end{eqnarray}
where $M_{\lambda,0}$ is the unreddened magnitude, $\omega_{\lambda}$ is the albedo of the dust grains and
the effective optical depth $\tau_{eff}$ is given by
\begin{equation}
\tau_{eff} = 2\int \sqrt{1-\omega_{\lambda}}\pi a^2Q_E(\lambda)n_dds
\end{equation}
where $a$ is the mean radius of the dust grains, supposed to be spherical, $Q_E(\lambda)$ is the extinction
(absorption+scattering) efficiency and $n_d$ is the dust number density. In our calculations, we have
assumed that the extinction efficiency and the albedo of the grains have the same properties as those of
galactic dust taken from Cardelli et al. (1989) and Li \& Draine (2001) respectively.

\section{Results}

If previous evolutionary models such as, for instance, those of Kodama \& Arimoto (1997) and IMP03 constrained
the free parameters by fitting the CMD of ellipticals, these models were unable to adequately reproduce 
the trend either in the $H\beta-Mg_2$ or in the $Mg_2-<Fe>$ diagrams. The behavior of our best models
in these diagrams is illustrated in fig.~1 (middle and lower panels). Indices 
for the same object measured by 
different authors, in spite of have being transformed into the Lick system, may differ
significantly among them which, besides the intrinsic (cosmic) dispersion, explains the scatter of 
data points in these diagrams. Even so, the trend of data points in both diagrams is now well
reproduced by our model sequence.

\begin{figure}

  \begin{center}
    \rotatebox{0}{ \includegraphics[height=6cm,width=8cm]{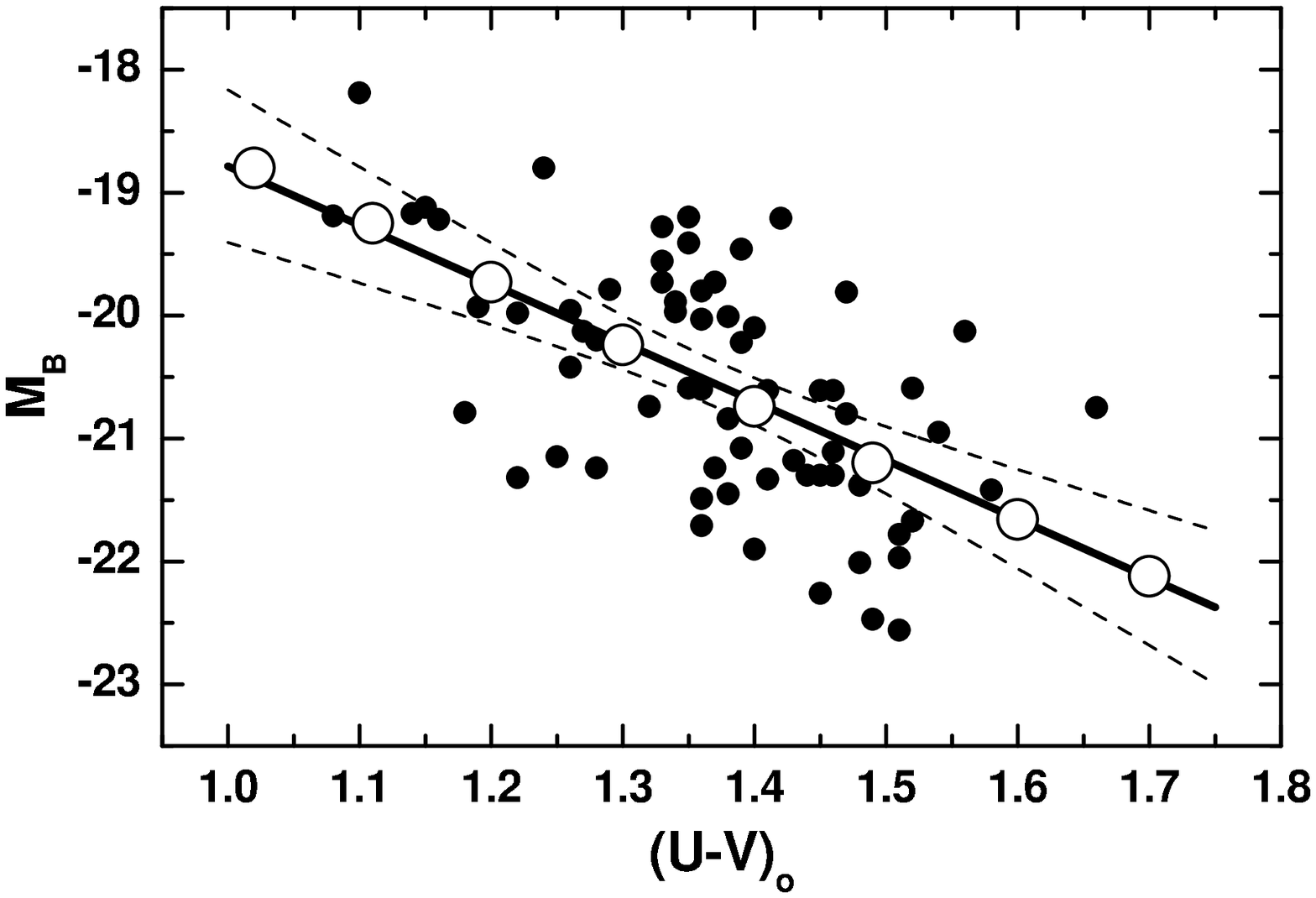}}
  \end{center}
  \vfill
  \vspace{-8mm}
  \begin{center}
    \rotatebox{0}{ \includegraphics[height=6cm,width=8cm]{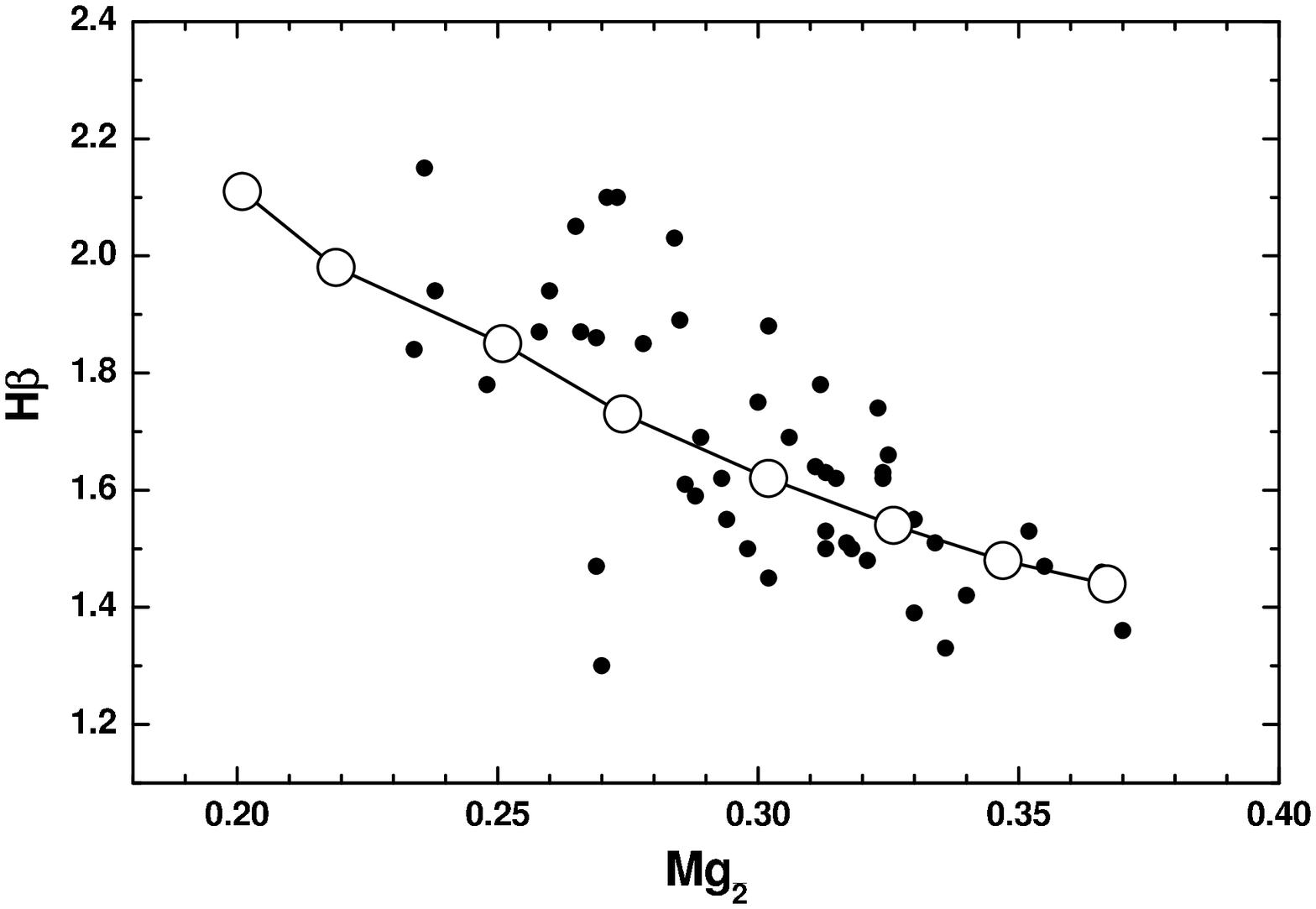}}
  \end{center}
  \vspace{-8mm}
  \begin{center}
    \rotatebox{0}{ \includegraphics[height=6cm,width=8cm]{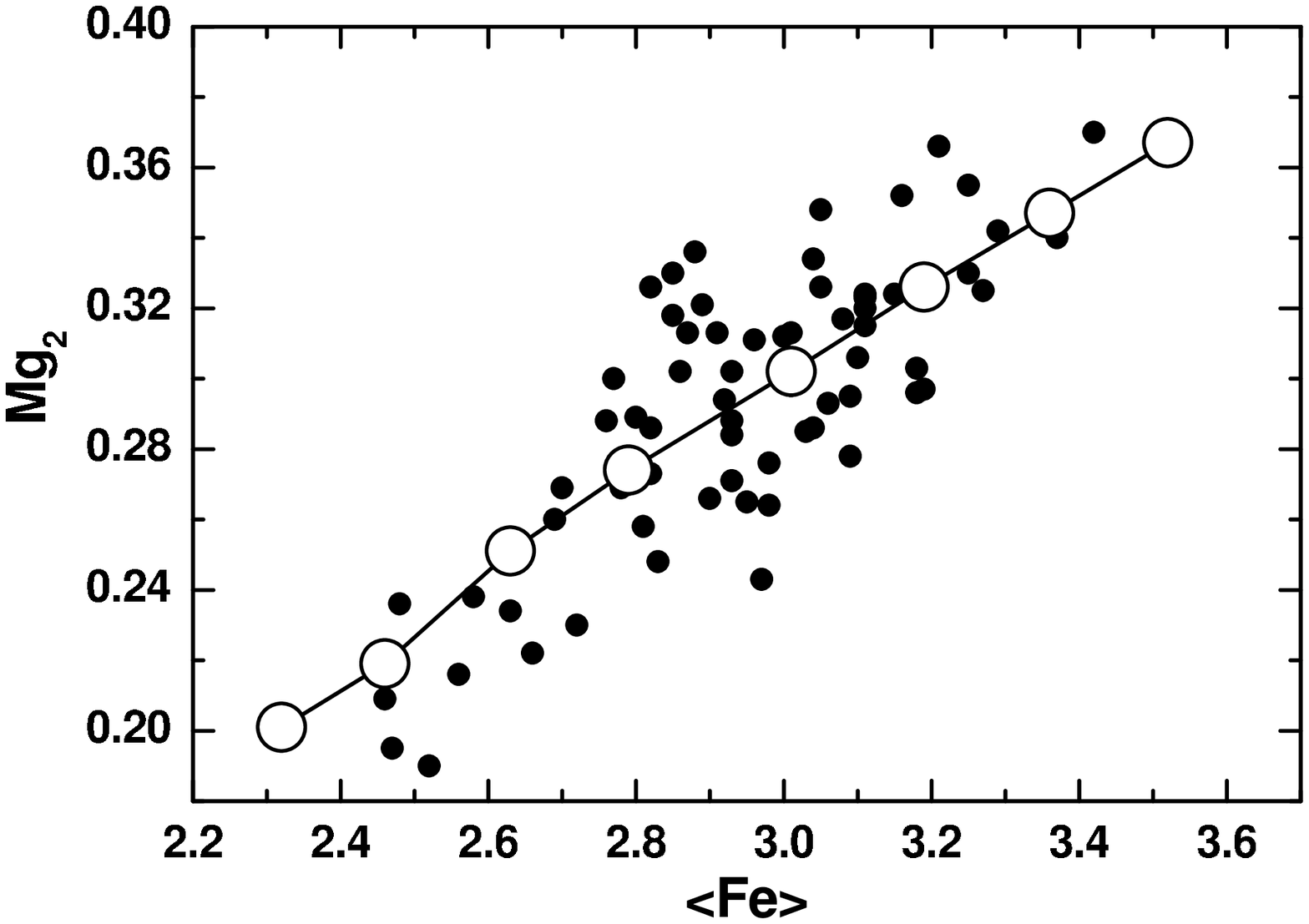}}
  \end{center}
\caption
{
The  observed CMD (solid points ) is shown in the upper panel, together with
the present theoretical results (open circles linked by black segments). The  
dashed lines indicate the mean dispersion. In the middle and in the lower panels, theoretical 
metallicity indices (open circles) are compared with data (solid points).  
Data are from: Worthey et al. (1992), Trager et al. (2000),
Mehlert et al. (1998; 2000; 2003), Kuntschner et al. (2001), Proctor et al. (2004)
and Howell (2005).}
\label{fig1}
\end{figure}

Table 1 gives the model parameters found to best describe the mean integrated properties of our
galaxies. The first column identifies the model and the second gives the mass parameter $M_0$ in
units of $10^{11} M_{\odot}$. The actual mass of the galaxy is $M_0(1-f_w)$, where the mass fraction
$f_w$ lost by the galaxy through the wind is given in seventh column. As expected, because of 
large potential wells, the mass 
loss fraction decreases for more massive galaxies. The star formation efficiency is given in
the sixth column and, as in IMP03, it increases for higher masses. A measure of the
down-sizing effect introduced in our models is the redshift $z_{80}$ at which 80\% of the mass of 
the galaxy was already assembled, given in the fifth column. We emphasize again that down-sizing was
introduced to improve the location of our models in the $H\beta-Mg_2$ diagram and it is
interesting to remark that the required assembling epoch for different masses is 
not in disagreement with recent observations mentioned above.    

\begin{table}
\caption{{\bf Model parameters:} 
columns give respectively the model identification (1), 
the mass parameter $M_0$ in units of $10^{11}$ M$_{\odot}$ (2), the 
infall time scale in Gyr (3), the wind time scale in Gyr (4), the redshift at which 80\% of the mass was 
assembled (5), the star formation efficiency in Gyr$^{-1}$ (6), the mass fraction lost by the wind (7) and
the exponent of the IMF (8)}
\label{tb1}
\centering
\begin{tabular}{@{}clllllll@{}}
\hline\hline
Model&$M_0$ & $t_{in}$ &$t_w$&$z_{80}$& k &$f_w$&$\gamma$\\
\hline
1&0.22&0.1&0.33&0.73&0.73&0.190&2.43\\
2&0.40&0.1&0.51&0.89&0.77&0.165&2.39\\
3&0.66&0.5&0.75&0.90&0.93&0.150&2.30\\
4&1.30&1.0&1.23&1.07&1.00&0.130&2.30\\
5&2.50&1.0&1.66&1.38&1.33&0.116&2.26\\
6&4.50&1.0&2.36&1.82&1.67&0.105&2.23\\
7&8.50&1.0&3.53&2.52&2.00&0.092&2.20\\
8&16.0&1.0&5.23&3.87&2.33&0.080&2.17\\
\hline
\end{tabular}
\end{table}

Table 2 gives some resulting model properties, including  the residual gas 
fraction, the mean stellar metallicity,
and the mean population age.

\begin{table}
\caption{{\bf Model properties:}
columns give respectively the model identification (1), the residual gas fraction (2), the fraction of hot gas (3),
the mean iron abundance (4), the mean stellar magnesium-to-iron ratio (5), 
the mean metallicity (6) and the mean population age
in Gyr (7)}
\label{tb2}
\centering
\begin{tabular}{c@{\hspace{1pt}}cl@{\hspace{8pt}}r@{\hspace{8pt}}c@{\hspace{8pt}}cl}
\hline\hline
Model&$f_g$ & $x_{h}$ &$[Fe/H]$&$[Mg/Fe]$&$[Z/H]$& age\\
\hline
1&0.0059&0.316&-0.204&+0.10&-0.24&6.4\\
2&0.0062&0.369&-0.139&+0.15&-0.16&7.2\\
3&0.0067&0.439&-0.016&+0.26&+0.02&7.3\\
4&0.0063&0.536&+0.002&+0.26&+0.02&8.6\\
5&0.0056&0.685&+0.082&+0.32&+0.12&9.7\\
6&0.0061&0.767&+0.157&+0.37&+0.20&10.8\\
7&0.0073&0.821&+0.260&+0.42&+0.43&11.8\\
8&0.0089&0.862&+0.437&+0.50&+0.47&12.8\\
\hline
\end{tabular}
\end{table}

Inspection of table 2 reveals some interesting characteristics of our models. Firstly, the fraction of
hot gas increases for more massive galaxies, as X-ray observations indicate, while for galaxies fainter
than $M_B \sim -20.0$ most of the residual gas is in the cold phase. For galaxies with absolute
magnitudes in the range $-20.3 < M_B < -18.5$, the amount of cold gas ($HI$) expected from our models
is given by the expression 
\begin{equation}
log\frac{M_{HI}}{M_{\odot}} = -0.677-0.455~M_B 
\end{equation}
The HI contents for a sample of E-galaxies in this luminosity range was measured by Lake \& Schommer
(1984) and are quite consistent with the amount of cold gas predicted by the above relation,
differing on the average by $\Delta log(M_{HI}/M_{\odot} \simeq$ 0.20.
Iron abundances range from $\sim$ half
of the solar value for faint objects up to three times solar for the brightest galaxies. The $[Mg/Fe]$ ratio
increases with the mass of the galaxy, consequence of an increasing star formation efficiency and a
slightly flatter IMF for massive galaxies, with both effects favoring the enrichment of the
ISM in $\alpha$-elements by type II supernovae. Mean population ages now span a  wider
range ($\sim$ 6 Gyr) than obtained in models without down-sizing. Notice that the mean
population ages given in table 2 are weighted by mass of alive stars. Had we used a luminosity weighted mean,
the resulting values would be reduced approximately by 0.3 Gyr. Larger differences would be expected in the
case of IMFs flatter than those characterizing the present models. In order to estimate
dust-to-gas mass ratios from the resulting residual gas fraction and the effective extinction, some
assumptions are required besides those already mentioned. These concern the average grain radius and
its intrinsic density. For this purpose, we have assumed
$a = 0.2\mu$m and $\delta = 1.5~gcm^{-3}$. Under these conditions, the dust-to-gas ratio varies
from $\sim$ 0.003 for the less massive galaxy up to 0.02, for the most massive modeled objects. These
values indicate that the relative amount of dust is compatible with the derived metallicities or,
in other words, with the amount of metals expected to be in the solid phase.

Tables 3 and 4 give the resulting spectrophotometric properties as colors, the absolute B-magnitude, the 
stellar mass-to-luminosity ratio, the Lick indices $H\beta$, $Mg_2$, $<Fe>$ and the dust effective extinction
(in mag) in the V band.

\begin{table}
\caption{{\bf Photometric properties I:}
columns give respectively the model identification (1), the absolute B-magnitude (2), the stellar 
mass-to-luminosity ratio (3), the dust effective extinction (in mag) 
(4) and the Lick indices $H\beta$, $Mg_2$ and $<Fe>$
respectively in columns (5), (6) and (7)}
\label{tb3}
\centering
\small
\begin{tabular}{c@{\hspace{4pt}}ccccc@{\hspace{4pt}}c}
\hline\hline
Model & $M_B$ &$M_*/L_B $&$A_V$&$H\beta$&$Mg_2$&$<Fe>$\\
\hline
1&-18.80&3.50&0.020&2.11&0.201&2.32\\
2&-19.25&4.32&0.058&1.98&0.219&2.46\\
3&-19.73&4.67&0.041&1.85&0.251&2.63\\
4&-20.24&5.90&0.062&1.73&0.274&2.79\\
5&-20.74&7.25&0.062&1.62&0.302&3.01\\
6&-21.20&8.64&0.080&1.54&0.326&3.19\\
7&-21.66&10.85&0.150&1.48&0.347&3.36\\
8&-22.12&13.54&0.201&1.44&0.367&3.52\\
\hline
\end{tabular}
\end{table}

\begin{table}
\caption{{\bf Photometric properties II:}
columns give respectively the model identification (1) and the integrated colors}
\label{tb4}
\centering
\begin{tabular}{clllll}
\hline\hline
Model &(U-V)&(B-V)&(V-R)&(V-I)&(V-K)\\
\hline
1&1.02&0.798&0.518&1.049&2.566\\
2&1.11&0.838&0.538&1.085&2.643\\
3&1.20&0.870&0.544&1.094&2.678\\
4&1.30&0.908&0.563&1.129&2.744\\
5&1.40&0.941&0.580&1.157&2.815\\
6&1.49&0.977&0.595&1.189&2.892\\
7&1.60&1.026&0.622&1.244&3.015\\
8&1.70&1.058&0.644&1.290&3.116\\
\hline
\end{tabular}
\end{table}

Using the derived properties of our models, we have derived relations between the mean age of
stellar population mix and observed integrated properties such as the indices $H\beta$, $<Fe>$ and
the color (U-V). These are
\begin{eqnarray}
\tau_1 = -19.230 + 3.884H\beta + 7.516<Fe>
\\
\tau_2 = 3.122 - 3.243(U-V) + 1.226<Fe>^2
\end{eqnarray}
where ages are in Gyr. The mean iron abundance and the mean magnesium-to-iron ratio
of the stellar population mix can be estimated from the relations
\begin{equation}
[Fe/H] = -1.694 - 3.220Mg_2 + 0.963<Fe>
\end{equation}
and
\begin{equation}
[Mg/Fe] = -0.674 + 0.363<Fe> - 0.318Mg_2
\end{equation}
Once the mean iron abundance and the mean magnesium-to-iron ratio are calculated from
the relations above, the mean metallicity can be derived from
\begin{equation}
[Z/H] = -0.313 + 0.399[Fe/H] + 1.242[Mg/Fe]
\end{equation}
 The scatter of data observed in figure 1 is due essentially to measurement errors and to
intrinsic variations of physical parameters among galaxies of same mass. We have estimated the
dispersion of data with respect to our models, correcting for measurement errors. Then, from
simple error propagation, we have estimated the uncertainties in the calibrations above. These
correspond to about 1.5 and 1.0 Gyr for ages estimated from eqs.~8 and 9 respectively, 0.17dex
for metallicities and 0.1dex for the [Mg/Fe] ratio. Systematic errors may certainly increase
these estimates.

In order to perform an application of the present models and compare with
the results derived by using the SSP approach, we have selected 42 galaxies
whose ages and metallicities were estimated by Howell (2005). For some of these galaxies,
Thomas et al. (2005) have also derived ages and metallicities, which will also be
used in our comparative analysis. Mean population ages and mean metallicities
derived from our calibrations are given in table 5. Ages are the average value 
resulting from the aforementioned calibrations and are followed by an 
index $a$ or $b$. The former implies that
the average age value differs by less than 1~Gyr from the individual determinations
whereas the later means that the difference is in the range 1-2~Gyr. Age differences
in this range implies that data points are not close to the mean theoretical
relations and, consequently, the resulting age or metallicity values are
more uncertain.

 \begin{table}
\caption{{\bf Ages \& Metallicities:}
 column (1) identifies the galaxy; column (2) gives the mean population age in Gyr;
the label (a) means that the average between the two calibrations differs by
less than 1~Gyr from individual age determinations while (b) means that the
difference is in the range 1-2~Gyr; columns (3) and (4) give the mean magnesium-to-iron
ratio and the mean metallicity, both with respect to solar values,  while the
next two columns, for comparison, give ages and metallicities derived 
from SSP by Howell (2005).}
\label{tb4}
\centering
\begin{tabular}{llllll}
\hline\hline
NGC&age&$[Mg/Fe]$&$[Z/H]$&age$_H$&$[Z/H]_H$\\
\hline
0315&10.7~(a)&+0.35&+0.23&5.0&+0.44\\
0584&11.3~(a)&+0.35&+0.26&2.4&+0.61\\
0596&8.2~(a)&+0.22&+0.26&4.4&+0.22\\
0636&10.5~(a)&+0.33&+0.23&3.8&+0.44\\
1052&8.3~(a)&+0.27&+0.03&16.0&+0.42\\
1172&7.8~(a)&+0.18&-0.07&4.8&+0.13\\
1209&9.3~(a)&+0.33&+0.17&15.6&+0.28\\
1400&8.1~(a)&+0.26&+0.02&14.2&+0.31\\
1700&12.0~(a)&+0.38&+0.33&2.3&+0.63\\
2300&10.3~(a)&+0.36&+0.22&5.5&+0.48\\
2768&8.2~(a)&+0.28&+0.18&10.0&+0.14\\
2778&8.8~(a)&+0.29&+0.12&5.0&+0.40\\
3115&11.7~(a)&+0.41&+0.36&3.9&+0.65\\
3193&8.3~(a)&+0.26&+0.02&11.8&+0.20\\
3377&7.8~(a)&+0.20&-0.02&3.5&+0.30\\
3379&10.3~(a)&+0.35&+0.23&8.0&+0.32\\
3607&8.6~(a)&+0.27&+0.05&10.6&+0.27\\
3608&7.9~(a)&+0.23&+0.00&6.1&+0.38\\
3610&10.8~(b)&+0.31&+0.00&1.7&+0.76\\
3640&9.4~(a)&+0.29&+0.15&4.9&+0.26\\
4168&8.9~(a)&+0.26&+0.09&5.0&+0.24\\
4365&11.3~(a)&+0.40&+0.36&5.9&+0.59\\
4374&10.0~(a)&+0.34&+0.22&11.1&+0.24\\
4472&11.9~(a)&+0.44&+0.42&7.8&+0.36\\
4473&10.5~(a)&+0.35&+0.24&4.0&+0.46\\
4486&8.8~(b)&+0.32&+0.14&19.6&+0.27\\
4489&9.1~(a)&+0.22&+0.14&2.3&+0.24\\
4552&11.1~(a)&+0.39&+0.29&10.5&+0.32\\
4621&10.5~(a)&+0.36&+0.22&15.8&+0.29\\
4697&8.4~(a)&+0.23&-0.02&7.1&+0.19\\
5576&10.2~(a)&+0.30&+0.17&2.5&+0.60\\
5638&8.6~(a)&+0.27&+0.05&7.8&+0.32\\
5813&9.3~(a)&+0.30&+0.12&14.9&+0.07\\
5831&7.6~(a)&+0.25&+0.04&2.7&+0.61\\
5846&8.3~(a)&+0.27&+0.06&12.2&+0.25\\
6702&9.2~(b)&+0.24&+0.07&1.4&+0.80\\
6703&9.6~(a)&+0.29&+0.11&3.9&+0.39\\
7454&7.3~(a)&+0.15&-0.15&4.7&+0.04\\
7562&10.2~(a)&+0.35&+0.25&7.1&+0.31\\
7619&12.2~(a)&+0.45&+0.41&13.5&+0.31\\
7626&10.7~(a)&+0.37&+0.24&12.0&+0.27\\
7785&10.6~(a)&+0.36&+0.26&7.9&+0.31\\
\hline
\end{tabular}
\end{table}


It is important to mention that ages derived from SSP models either by Howell (2005)
or by Thomas et al. (2005) are consistent but with values given by the latter authors
being systematically higher by $\sim$ 1.5 Gyr than those by the former author. 
The same remark is valid for metallicities, excepting that
for $[M/H] > 0.5$ the values derived by Howell are higher than 
those by Thomas et al. In fig.~2 (upper panel) we
have plotted metallicities as a function of ages derived from SSP, using the
results by Howell and including those by Thomas et al. for the same objects.
In spite of the significant scatter, we notice that "young" 
objects, with ages $<$ 5~Gyr are those with the highest metallicity while galaxies
with lower metallicities are generally old. This behavior is completely different
when a similar plot is performed with values obtained from our evolutionary
models (fig.~2, lower panel). In this case, a robust correlation between
the mean metallicity and the mean population age is observed, indicating that
the most massive galaxies are those which have been more enriched in chemical
elements thanks to the higher star formation efficiency. The most discrepant object
in this plot is NGC 3610, for which the mean population age determination gives 
values discrepant by $\sim$ 2.5 Gyr, indicating that the model parameters which characterize
this galaxy are far from those describing the mean trend. This galaxy is likely
to be a merger remnant, and merits more detailed structural modeling (cf. Strader, Brodie and Forbes 2004),
beyond the scope of this paper. However, if an important merger event stimulates the star 
formation activity, this can roughly be simulated in our approach by a higher star formation 
efficiency. In fact, with $k = 3.96~Gyr^{-1}$ (instead of $1.33~Gyr^{-1}$ for a galaxy of same 
luminosity) it is possible to reproduce quite well the observed colors and Lick indices of NGC 3610.
In this case, the mean population age would be $2.8~Gyrs$, indicative of the epoch of the
merger episode.  Notice also that, in general, our derived ages span the 
interval 7-12 Gyr while for the same galaxies, the
resulting interval from the SSP approach is 2.3-19 Gyr.

\begin{figure}
\begin{center}
\rotatebox{0}{ \includegraphics[height=6cm,width=8cm]{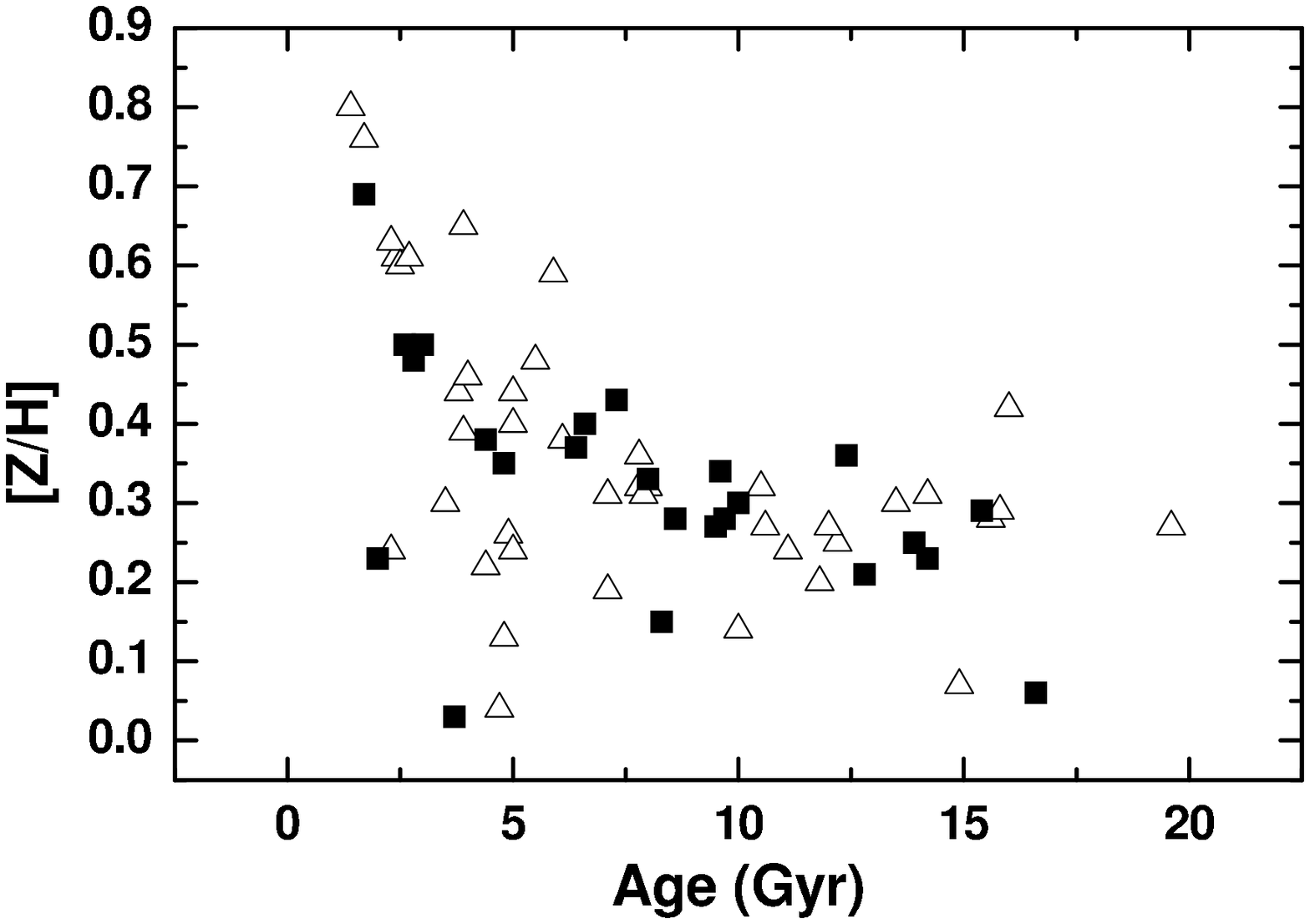}}
\end{center}
\vfill
\vspace{-8mm}
\begin{center}
\rotatebox{0}{ \includegraphics[height=6cm,width=8cm]{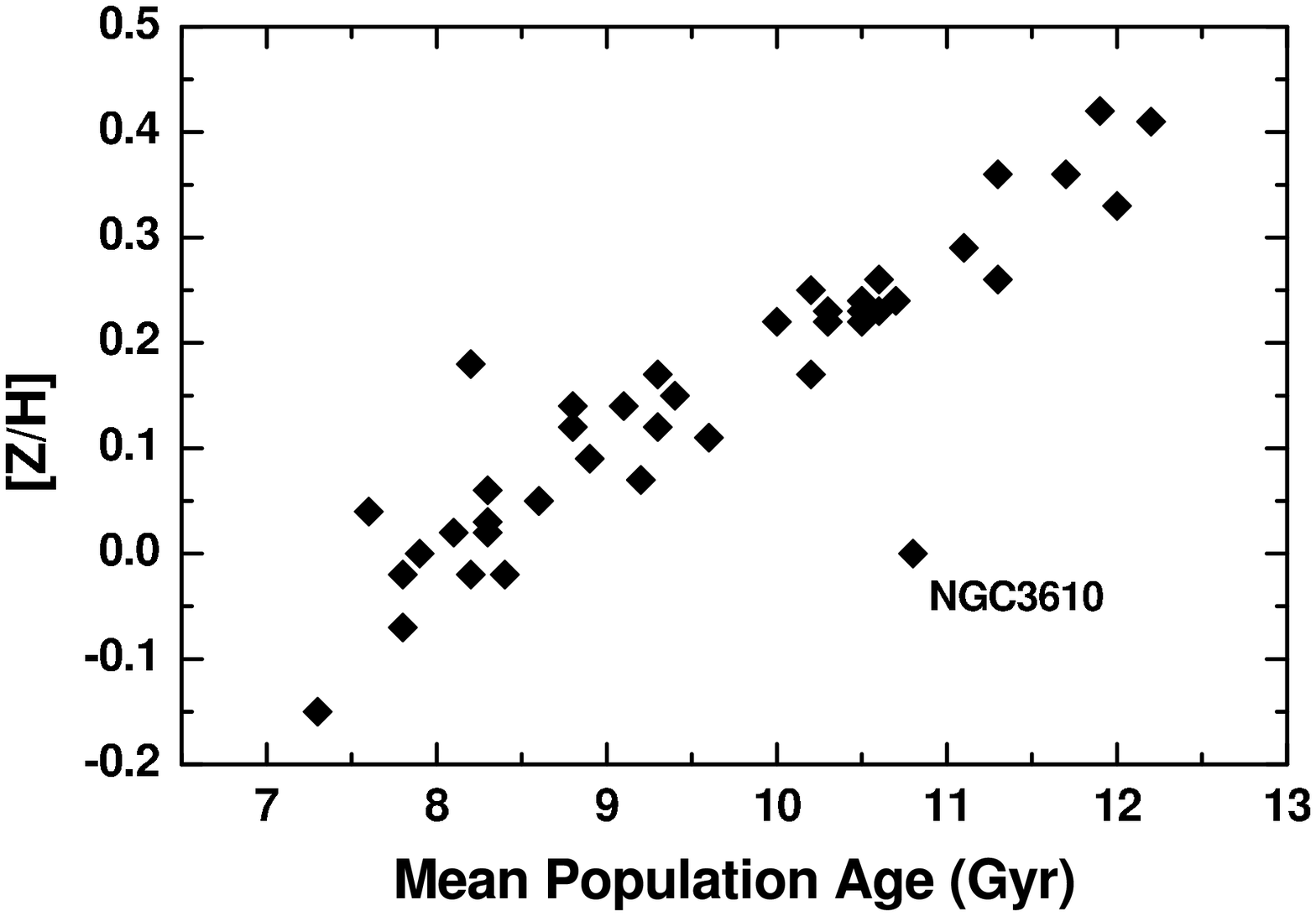}}
\end{center}
\caption{Upper panel: metallicities as a function of ages derived from SSP models. Open triangles
are Howell (2005) data while solid squares are Thomas et al. (2005) data for common
galaxies listed in table 4. Lower panel: the same
plot but for ages and metallicities derived from the present model.}
\label{fig2}
\end{figure}

It is worth mentioning a further aspect when comparing results derived from SSP models and EVM.
The present sequence of models was built by varying different parameters in order to
adequately reproduce the integrated properties such as luminosities, colors and metallicity indices. In this
sense, the present models are ``self-consistent". In general, analyses based on SSP models
use only metallicity indices but not colors. If ages and metallicities derived from SSP models
are used to predict the colors of the object, we obtain values considerably redder in
comparison with observations. The reason for this behavior
is that all stars in SSP models have the same metallicity while in
evolutionary models, stars of low-metallicity are also present in the population
mix, contributing to the total light and producing bluer colors. This effect is
illustrated in fig.~3 in which, for galaxies listed in table~5, the predicted (U-V) colors
derived from SSP metallicities and ages are plotted against observed values. 
 
\begin{figure}
\centerline{
\vspace{-0.8cm}
\epsfxsize=0.5\textwidth\rotatebox{0}
    {\epsfbox{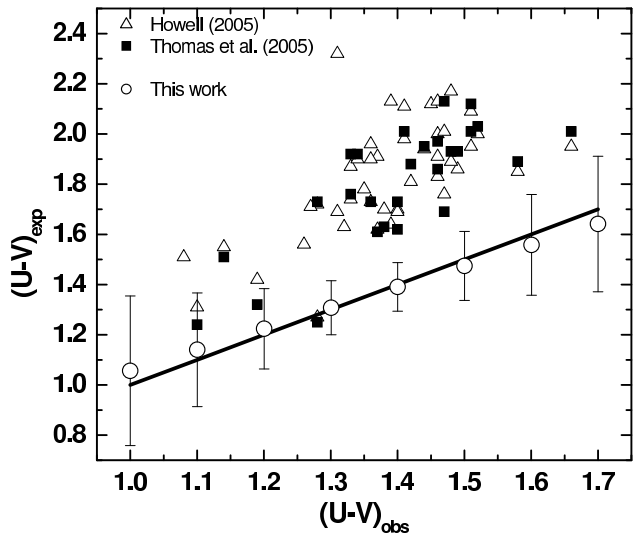}}
           }
\vspace{0.8cm}
\caption{Predicted versus observed (U-V) colors. The open triangles and solid squares 
identify the source for SSP ages and metallicities. The solid line represents the expected
correspondence. Open circles represents predictions of our models. Observed colors were
averaged within bins of magnitude $\Delta~M_B$. The corresponding dispersions are also indicated.}
\label{fig3}
\end{figure}

\section{Conclusions}

In this study, new models for E-galaxies are presented. The ``canonical" one-zone
model was revised by including a simplified two-phase interstellar medium, in
which stars are formed only in cold regions and a galactic wind removes hot gas
from the galaxy. The mass of the galaxy increases with time as a consequence of
infall, here represented by either continuous accretion or discrete minor merger events. The
possibility that some dust is mixed with the residual gas was also taken into
account. The different parameters of the model were varied in order 
to adequately reproduce integrated properties such as the color-magnitude diagram 
and the trends of indices such as  $H\beta$ vs. $Mg_2$ and $Mg_2$ vs. $<Fe>$. 

The fundamental characteristic of these models concerns the beginning of stellar
assembly and  star formation activity. The start-up time was adjusted
in order to give a better description of data in the $H\beta$-$Mg_2$ diagram
and this requires that the less massive galaxies are assembled later, in
agreement with the down-sizing effect. In order to characterize this effect,
we have defined the quantity $z_{80}$, corresponding to the redshift at which
80\% of the (baryonic) mass was assembled. Galaxies in the luminosity
range $-18.8 > M_B > -20.2$ were assembled in the redshift range $0.7 < z_{80} < 1.1$
whereas brighter objects $-21.2 > M_B > -22.2$ were assembled earlier, e.g.,
$1.8 < z_{80} < 3.9$.

Our models predict an increasing fraction of hot gas for massive galaxies, a consequence
of gravitational trapping, in concordance with X-ray observations. The
derived amount of dust mixed with the residual gas also increases with galaxian
masses and the dust-to-gas ratio varies from 0.003 for faint up to 0.02 for bright
galaxies. However, these numbers depend on the adopted grain dimension and composition.
Thanks to down-sizing, the resulting mean stellar population ages span a range 
of 6.4-12.8 Gyr, considerably wider than our previous models (IMP03). Metallicities vary
from half to about three times solar, increasing with the mass of the galaxy. These
results imply that the CMD is not either a pure metallicity or a pure age sequence, but
a combination of these two quantities.

From the present models, different calibrations were obtained allowing estimates of
mean properties of the stellar population mix like the age, the metallicity and the
$[Mg/Fe]$ ratio, from the knowledge of integrated parameters like the $(U-V)$ color and
the Lick indices $H\beta$, $Mg_2$ and $<Fe>$. These calibrations were applied to
a sample of early type galaxies, whose spectral indices were interpreted 
in terms of SSP models. From these studies based on SSP models, galaxies
with stellar population ages as young as 2.5~Gyr can be found, while the corresponding ages
derived from our model calibrations are significantly higher, ranging from 7.5 up to 12 Gyr. 

The mean values for metallicities and ages concern the bulk of the stellar population mix
constituting ellipticals. However, the residual gas originated either from the stellar evolution
or infall can presently be converted into stars. In fact, Ferreras \& Silk (2000) studied a
sample of early type galaxies in Abell 851 and found that the slope and scatter in the
near-ultraviolet (NUV)-optical plane are consistent with some objects having $\sim$ 10\%
of their stellar mass in stars younger than $\sim$ 0.5 Gyr. The study of a large sample
by Kaviraj et al. (2006) confirms such a conclusion. They have noticed that about 30\%
of the galaxies in their sample ($\sim$ 2100 objects) have UV-optical colors consistent
with some star formation activity within the last Gyr. This recent activity represents, according
to their estimates, about 1-3 \% of the stellar mass in stars less than 1 Gyr old. Using
the fraction of residual cold gas and the star formation efficiency of our models, we have
estimated the mass fraction of stars formed in the last 0.8 Gyr, which are in the range
2.3-5.7~\%, fully consistent with UV and NUV observations.

As mentioned, the present models describe mean integrated properties and some
uncertainties are certainly present in the resulting calibrations of age and metallicity,
because of cosmic scatter and observational errors. In a future paper we will
report the application of the present model to individual galaxies, deriving the
relevant parameters from the best fit of the integrated parameters of each object.
This approach will possibly give information about the cosmic scatter and
will eventually reveal any possible correlations with the environment. 
 
\vspace{0.5cm} 

\noindent
{\bf Acknowledgments}
T.P. Idiart thanks to FAPESP (grant 2006/01025-3) for the financial support
during the stay at Oxford.

\noindent


\end{document}